%% file: condmat1130.tex
\documentclass[aps,pra,twocolumn,amsmath,amssymb,superscriptaddress,floatfix]{revtex4-1}

\usepackage{graphicx}
\usepackage{multirow}
\usepackage{color}
\usepackage{bm}

\renewenvironment{acknowledgments}[1]{%
 {\textit{Acknowledgment ---} {#1}}%
}

\def\infinity{\infty}
\def\t#1{\textrm{#1}}
\def\ket#1{|#1\rangle }
\def\bra#1{\langle #1 |}
\def\braket#1{\langle #1 \rangle}
\def\n{\nonumber \\ }

\begin{document}

\title{Current-voltage characteristic and shot noise of shift current photovoltaics
}

\author{Takahiro Morimoto}
\affiliation{Department of Physics, University of California, Berkeley, CA 94720, USA}

\author{Masao Nakamura}
\affiliation{RIKEN Center for Emergent Matter Science (CEMS), Wako, Saitama, 351-0198, Japan}
\affiliation{PRESTO, Japan Science and Technology Agency (JST), Kawaguchi, Saitama, 332-0012, Japan}

\author{Masashi Kawasaki}
\affiliation{RIKEN Center for Emergent Matter Science (CEMS), Wako, Saitama, 351-0198, Japan}
\affiliation{Department of Applied Physics and Quantum-Phase Electronics Center, University of Tokyo, Bunkyo, Tokyo 113-8656, Japan}

\author{Naoto Nagaosa}
\affiliation{RIKEN Center for Emergent Matter Science (CEMS), Wako, Saitama, 351-0198, Japan}
\affiliation{Department of Applied Physics and Quantum-Phase Electronics Center, University of Tokyo, Bunkyo, Tokyo 113-8656, Japan}

\date{\today}

\begin{abstract}
We theoretically study the current-voltage  
relation, $I-V$ characteristic, of the photovoltaics due to the shift
current, i.e., the photocurrent generated 
{\it without} the external dc electric field in noncentrosymmetric crystals
through the Berry connection of the Bloch wavefunctions.  
 We find that the $I-V$ characteristic and shot noise 
are controlled by the difference of group velocities between conduction and 
valence bands, i.e., $v_{11}-v_{22}$, and the relaxation time $\tau$.
Since the shift current itself is independent of these quantities, there are 
wide possibilities to design it to maximize the energy conversion rate and
also to suppress the noise. We propose that the Landau levels in 
noncentrosymmetric two-dimensional systems are the promising 
candidate for energy conversion. 
\end{abstract}

\maketitle

\textit{Introduction ---}
 There are variety of nonlinear optical processes which are intensively studied 
especially since the strong laser light became available \cite{Bloembergen,Boyd}. 
Of particular interest is the generation of dc photocurrent induced by 
the light excitations aiming at the application to the solar cells.
In the conventional setup, the light is injected to the p-n junction and 
the photo-generated electrons and holes are separated by the built-in 
potential gradient. In this situation, the diffusive motion
of electrons and holes into the electrode is required to obtain the 
dc current.  On the other hand, the focus of recent intensive attention is the
shift current in noncentrosymmetric crystals, which does not require the 
interface such as p-n junction 
\cite{Kraut,Sipe,Young-Rappe,Young-Zheng-Rappe,Cook17,Morimoto-Nagaosa16}. 
In this case, the broken spatial inversion symmetry 
$\mathcal{I}$ of the crystal structure determines the direction of the photocurrent 
{\it without } the external dc electric field. The shift current is regarded as one
of the possible microscopic mechanisms of the highly efficient solar cell action
in perovskite oxides \cite{Nie,Shi,deQuilettes,Bhatnagar}. From the theoretical point of view, shift current 
originates from the quantum geometric nature of the 
Bloch wavefunctions in noncentrosymmetric solids \cite{Morimoto-Nagaosa16,Nagaosa-Morimoto17}.
Due to the conjugate relation between the position $x$ and the 
momentum $p=\hbar k$, $x$ is represented by $i \hbar \partial/\partial p$.
($\hbar$ is the Planck constant divided by $2 \pi$.) From this relation applied to the Bloch wavefunctions of solids,
there appears the intracell coordinate $x_n$
for the wavepacket made from each Bloch wavefunction
$ \braket{ x|\psi_n (k) } = e^{ik x} \braket{ x|nk }$ as
\begin{equation}
x_n = a_n(k) = -i \braket{ n k | \nabla_k |nk }
\end{equation}
which is called Berry connection \cite{Resta}.
The optical transitions between the valence and conduction bands induces
the intracell coordinates to shift, which results in the dc current. This is the 
mechanism of the shift current in noncentrosymmetric crystals.
(Note that $a_n(k)$ is 0 for the centrosymmetric crystal with 
time-reversal symmetry.)
The expression for the shift current contains only the inter-band 
matrix elements of the current, which is related to the Berry connection defined above, 
in sharp contrast to the usual photocurrent by the classical motion of 
carriers described by the intra-band matrix elements of the current, i.e,
the group velocity. Therefore, it is expected that the physical nature of the
shift current is distinct from that of the conventional current.
  
In Ref.~\cite{Morimoto-Nagaosa16}, a theoretical framework has been developed to treat the
shift current of the two-band model analytically, by combining the
Keldysh method and Floquet formalism. The relaxation of electrons 
$\Gamma= \hbar/\tau$ is introduced, and the competition between the
stimulated transitions, i.e., recombining electrons and holes, 
and the non-radiative relaxation is found to be
described by the factor $\Gamma E^2/\sqrt{(e E r)^2+ \Gamma^2}$ ($r$ is a constant having a dimension of length).
Namely, the non-radiative relaxation $\Gamma$, 
which has no directional dependence, is needed for the finite 
shift current, while the stimulated transition induces the back flow of the current.
This dependence on the strength of the electric field $E$ has been 
recently observed in SISb by THz spectroscopy \cite{Sotome18}.

\begin{figure}
\includegraphics[width=1.0\linewidth]{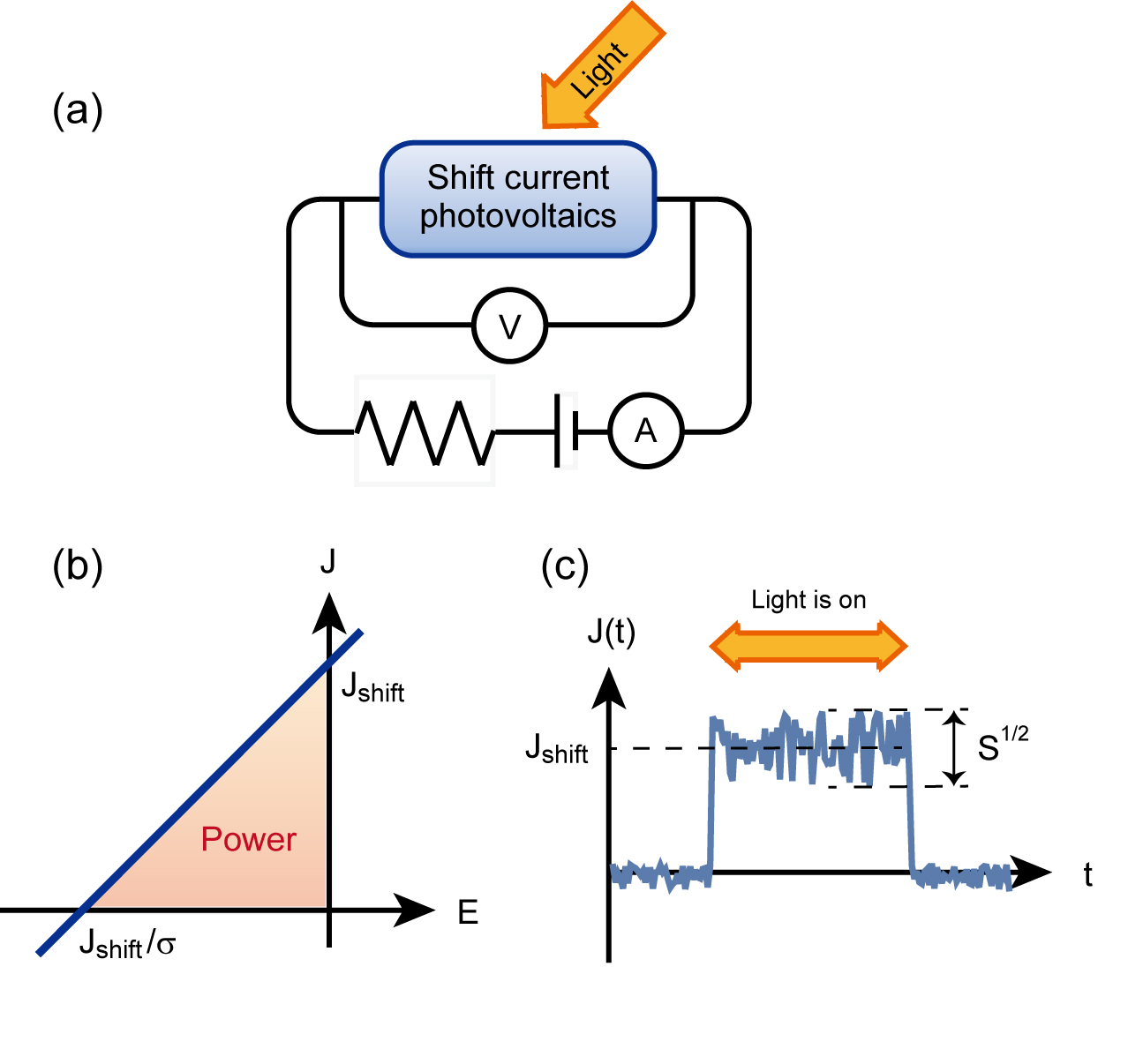}
\caption{\label{fig: schematics}
Schematic picture of $I-V$ characteristics and photodetection in shift current photovoltaics. 
(a) Setup to measure $I-V$ and photocurrent. 
(b) $I-V$ characteristics of shift current photovoltaics. In this work, we consider current density $J$ as a function of the electric field $E$.
(c) Shot noise in shift current photovoltaics. When the light is off, current fluctuates around zero due to thermal noise which can be neglected in zero temperature. 
When light is on, shift current flows and additional current fluctuation appears due to photocarriers in the nonequilibrium steady state.
}
\end{figure}

In realistic setup of solar cell action, the voltage at open circuit conditions is
often measured, which determines the power conversion rate. 
Therefore, the $I-V$ characteristics of the shift current is an important issue [Fig.~\ref{fig: schematics}(a, b)].
Another possible application is the photodetector [Fig.~\ref{fig: schematics}(c)]. For that purpose, 
the noise of the shift current needs to be analyzed.  
In the present paper,
we theoretically study these two issues.
The former is achieved by developing a new theoretical technique that combines gauge invariant formulation of Keldysh Green's function with Floquet theory description of shift current~\cite{Morimoto-Nagaosa16}.
The latter is done by computing correlation functions with nonequilibrium Green's functions on top of Floquet theory approach.

\textit{$I-V$ characteristics of shift current photovoltaics ---}
 In the experimental setup to measure shift current, an electrical circuit is formed by attaching two electrodes to the crystal and including resistors as shown in Fig.~\ref{fig: schematics}(a). 
The relationship between the voltage 
between the two electrodes and the
current flowing through the crystal is the fundamental information that is necessary to 
design the solar cells [Fig.~\ref{fig: schematics}(b)]. 
Especially, the open circuit voltage 
is an important quantity for the energy conversion rate.
The shift current involves the inter-band transitions
which inevitably create the electrons and holes. Therefore,
the additional current due to these photogenerated carriers
and their interference with the shift current should be analyzed,
which we undertake in this section.

We study a system in an external electric field by using gauge invariant formulation of Keldysh Green's function 
and its gradient expansion.
In the presence of an external dc electric field $E_\t{dc}$, the Green's function is
expanded with respect to $E_\t{dc}$ as
$G (E_\t{dc}) = G_0 + (\hbar e E_\t{dc}/2) G_E +O(E_\t{dc}^2)$
\cite{Onoda06,Sugimoto08,Morimoto18}.
In addition, we employ Floquet two band model to incorporate ac electric field $E_\t{light}$ of constant light that is irradiated to the sample and produces shift current \cite{Morimoto-Nagaosa16}.
We focus on the valence band with one photon and the conduction band with zero photon.
The Floquet Hamiltonian is given by
\begin{align}
H_F&=
\begin{pmatrix}
\epsilon_1 + \hbar \omega & e A v_{12} \\
e A v_{21}  & \epsilon_2
\end{pmatrix},
\end{align}
when the static Bloch Hamiltonian $H$ (without the light field) is diagonalized as $H(k)\ket{u_i(k)}=\epsilon_i(k) \ket{u_i(k)}$ with 
 the energy dispersion $\epsilon_i(k)$ and the Bloch wave function $\ket{u_i(k)}$ for the $i$th band.
Here, the velocity matrix element is defined as $v_{ij}=(1/\hbar) \bra{u_i} \partial_k H \ket{u_j}$, and
$A=E_\t{light}/\omega$ with the strength $E_\t{light}$ and the frequency $\omega$ of the light electric field.
The indices 1 and 2 refer to valence and conduction bands, respectively. (For details, see Appendix \ref{app: floquet}.)

With this setup, the $I-V$ characteristic of shift current materials is given by
\begin{align}
J(E_\t{dc})&= J_\t{shift} + \sigma_E E_\t{dc},
\end{align}
with
\begin{align}
J_\t{shift}
&= \frac{2\pi e^3}{\hbar^2 \omega^2} |E(\omega)|^2
\int[dk] \t{Im} \left[ \left(\frac{\partial v}{\partial k} \right)_{12} v_{21} \right] \delta(\omega_{21}-\omega),
\label{eq: J shift}
\\
\sigma_E&= \frac{4 \pi e^4}{\hbar^3 \omega^2} |E(\omega)|^2 \tau^2 \int [dk]
|v_{12}|^2 (v_{11}-v_{22}) R' \delta(\omega_{21}-\omega),
\end{align}
where $[dk] \equiv dk/(2\pi)^d$ with the dimension $d$, 
$R=\t{Im}[(\partial_k v)_{12}/v_{12}]$,
$R'=\t{Re}[(\partial_k v)_{12}/v_{12}]$,
and $\omega_{21}=(\epsilon_2-\epsilon_1)/\hbar$.
We note that $R=\partial_k \t{Im}[\log v_{12}] + a_1 -a_2$ is called shift vector related to the shift of intracell coordinates between the two bands.
(For derivation, see Appendix \ref{app: I-V}.)

The physical meaning of the above expressions can be understood as follows. 
In our setup, the sample is subjected to constant light field, which produces the constant shift of the electrons associated with the inter-band transitions leading to the shift current $J_\t{shift}$. With the dc electric field, the accelerated motion of the photoexcited electrons and holes generates additional current $\sigma_E E_\t{dc}$ which is proportional to the difference of the group velocities between the conduction and valence bands, i.e., $v_{11}-v_{22}$. 
The former, i.e., shift current $J_\t{shift}$, is basically independent of the lifetime $\tau$ of the electrons, while the latter is proportional to $\tau^2$.
This is because the photocarrier density is proportional to the recombination time $\tau$ and their mobility is also proportional to the scattering time $\tau$. 
(Note that both recombination and scattering times are described by the same $\tau$ in our formalism.)
Therefore, it is expected that the shift current $J_\t{shift}$ (at zero bias) is almost independent of the disorder and temperature of the sample, while the slope in $I-V$ characteristics that corresponds to the additional current driven by $E_\t{dc}$ is strongly disorder and temperature dependent through the $\tau^2$ factor.

\textit{Current noise in shift current photovoltaics ---}
The transport current is inevitably associated with the noise. There are two types of the current noise \cite{Imry,Lifshitz-Pitaevskii}. 
One is the equilibrium noise, i.e., Nyquist noise, 
which is independent of the details of the system and is given only by the impedance 
$Z(\omega)$ of the current circuit at frequency $\omega$. The spectrum of the current fluctuation 
is given by $(J^2)_\omega = [ \omega \Re[ 1/Z(\omega) ]\coth(\omega/2T)$. The other is 
the nonequilibrium noise induced by the current flow. At low temperature, the quantum 
mechanical shot noise is the dominant contribution, which originates from the discrete nature of
the electron and its charge $-e$. The shot noise is characterized by the Fano factor 
$F=(J^2)_{\omega=0}/I$ with $I$ being the current. For the ideal situation, $F=2e$
and the shot noise is used as a tool to determine the charge of the carriers.

As mentioned in the introduction, the shift current is distinct from the conventional 
current in that shift current is intimately related to the wave nature of the electrons that is characterized by the
off-diagonal matrix elements of the current operator. 
Therefore, it is expected that the noise of the shift current is distinct from that of 
the conventional one.

Motivated by this consideration, we study current noise in shift current photovoltaics. We compute autocorrelation of local current operator $v_\t{loc}$ defined 
at one point (e.g. $x=0$) in the bulk, which is given by
$
v=\frac{1}{2 L} \sum_{k,k'} (v_k + v_{k'})c_k^\dagger c_{k'},
$
where $L$ is the sample size, and $k$ and $k'$ satisfy $|k-k'|<1/l$ with the size of the electrode $l$ \cite{Marel15}.
The current  noise is given by the zero frequency component of the autocorrelation as
\begin{align}
S &=\int dt (\braket{v_\t{loc}(t) v_\t{loc}(0)} - \braket{v_\t{loc}}^2).
\end{align}
To describe the nonequilibrium steady state under light irradiation that supports shift current, we again use Floquet two band model $H_F$.
By using Green's functions in Floquet two band model derived in Appendix \ref{app: floquet}, the noise $S$ can be expressed as
\begin{align}
S= \frac{e^4}{\hbar^2 \omega^2} E^2 \tau \int [dk]  |v_{11}-v_{22}| |v_{12}|^2 \delta(\omega_{21} - \omega).
\label{eq: S shift}
\end{align}
(For derivation, see Appendix \ref{app: noise}.)
This expression shows that the noise is proportional to the relaxation time $\tau$ and may be interpreted as temperature noise in 
the steady state with the effective temperature proportional to the transition rate 
$\approx |v_{12}|^2$.
Interestingly, the noise $S$ does not have a part that corresponds to usual current noise in the equilibrium systems which is proportional to the current ($S\propto J$).
This means that the noise $S$ is substantially suppressed if the relaxation time $\tau$ is very short or two bands are parallel with each other ($v_{11}-v_{22} \approx 0$),  In such cases, shift current 
photovoltaics can show much less noise than conventional metals subject to current noise.

\textit{Shift current in Landau levels ---}
In the previous sections, we have shown that both the 
$I-V$ characteristics and the noise are governed by the
difference of group velocities of conduction and 
valence bands, i.e., $v_{11}-v_{22}$, and the relaxation time $\tau$.
In particular, when $v_{11}-v_{22}=0$, the current is independent of the
voltage and also the noise is 0. This is an ideal situation both for 
the solar cells and photodetectors. However, it is usually difficult to 
realize the situation of ``parallel'' dispersion between the conduction and
valence bands. 
Here we propose the dispersonless Landau levels in 2D
offers a promising laboratory to test the ideal developed above. 
The dc current in the flat band system can be also a smoking gun experiment
for the unconventional nature of the shift current distinct from the
usual photocurrent by the diffusive motion of photocarriers. 
More explicitly, we consider the two systems, i.e., the
Landau levels of graphene and Landau levels of the surface 
state of three dimensional topological insulator.

\begin{figure}
\includegraphics[width=0.8\linewidth]{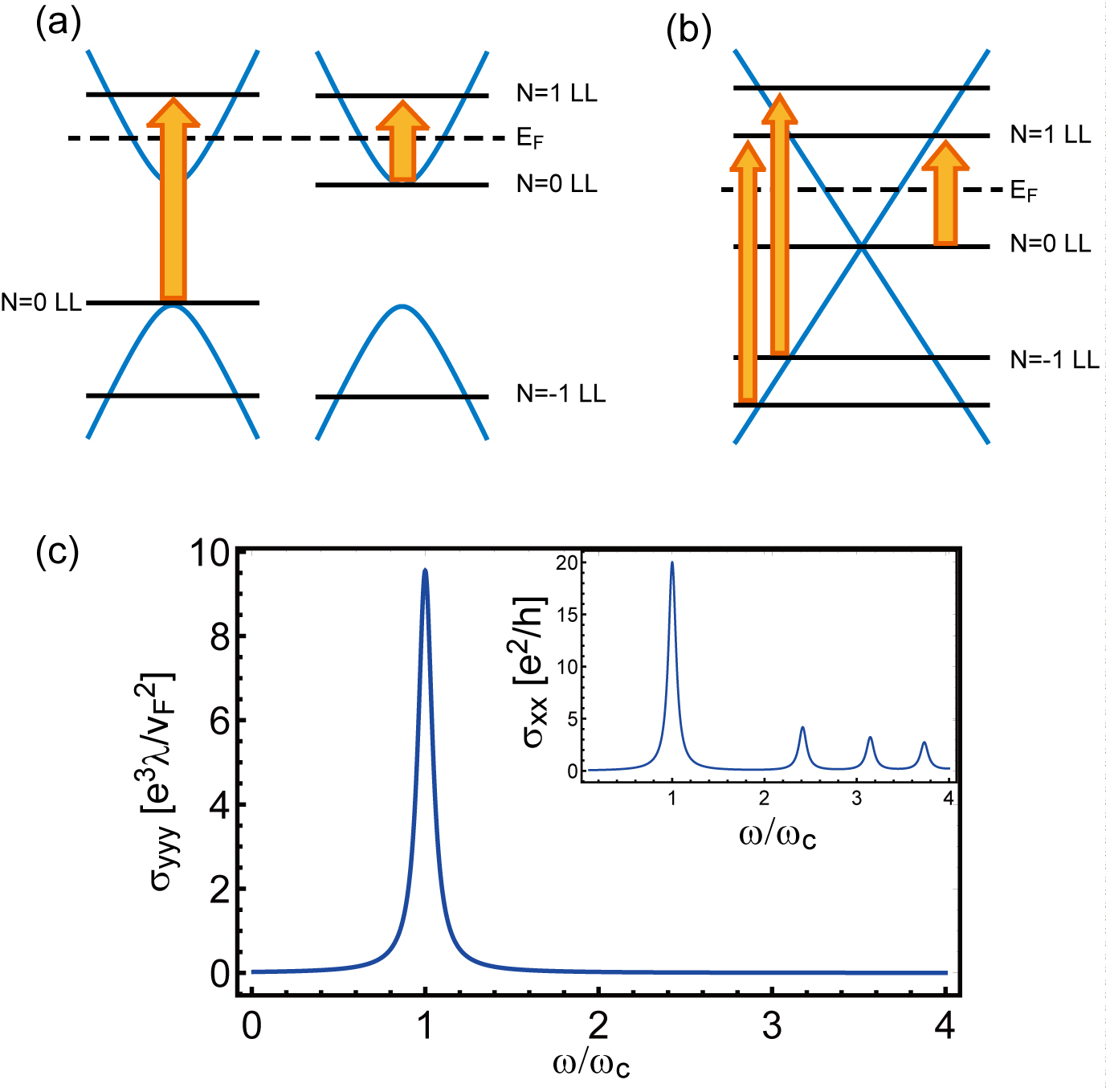}
\caption{\label{fig: LLs}
Optical transitions in LLs in inversion broken 2D systems realized in (a) graphene with staggered potential, and (b) TI surface states.
(c) Shift current in TI surface states. We plot the nonlinear conductivity $\sigma_{yyy}$ as a function of frequency. The inset is for linear conductivity $\sigma_{yy}(\omega)$. Relaxation time is set to $1/\tau=0.05 \omega_c$.
}
\end{figure}

We first study shift current in graphene Landau levels (LLs).
Since LLs have flat band dispersion, we expect that the current noise is suppressed and the power of shift current is enhanced in LLs.
In order to take into account the effect of inversion breaking in graphene,
we consider Dirac Hamiltonian with a trigonal warping term \cite{Ando98,Koshino-McCann09,McCann-Koshino13},
\begin{align}
H &= v_F
\begin{pmatrix}
0 & \pi^\dagger \\
\pi & 0 \\
\end{pmatrix}
+ \eta \kappa
\begin{pmatrix}
0 & \pi^2 \\
(\pi^\dagger)^2 & 0 \\
\end{pmatrix},
\end{align}
with Fermi velocity $v_F$, strength of trigonal warping $\kappa$,
$\pi=k_x \pm i k_y$, and $\eta=\pm 1$ for K and K' valleys, respectively.
In the presence of magnetic field,
the momentum operators are written as
$\pi=(\sqrt 2 \hbar/\ell)a^\dagger$ at K and 
$\pi=(\sqrt 2 \hbar/\ell)a$ and K' 
with annihilation and creation operators $a$ and $a^\dagger$, and the magnetic length $\ell=\sqrt{\hbar/eB}$. 
Now let us consider shift current supported by the graphene LLs. We suppose that the Fermi 
level is located between $n=0$ and $n=1$ LLs, and the sample is irradiated with linearly 
polarized light along the $x$ direction with the photon energy close to $\hbar \omega_c$.
In this case, the photoexcitation takes place from $n=0$ LL to $n=1$ LL. The shift current 
response $J_y = \sigma_{yxx} |E_x(\omega)|^2$ is given by
\begin{align}
\sigma_{yxx}(\omega)= \eta \frac{e^3}{\hbar \omega^2 \ell^2} v_F \kappa \delta(\omega-\omega_{10}),
\end{align}
where $\omega_{10}=(1/\hbar)(\epsilon_1 - \epsilon_0)$ with $\epsilon_0$ ($\epsilon_1$) being the energy of $n=0$ ($n=1$) LL.
This means that the shift current response vanishes when the optical transition $0 \to 1$ 
takes place both at K and K' valleys.
This is natural since the original Hamiltonian has inversion symmetry 
$\mathcal{I}=\sigma_x \tau_x$, where $\sigma$ and $\tau$ are Pauli matrices acting on 
sublattice and valley degrees of freedom;
nonzero shift current requires breaking of inversion symmetry. One way to achieve this is 
introducing staggered potential $m \sigma_z \tau_0$. For example, this situation is realized in graphenes on the substrate of BN, and also in transition metal dichalcogenide (TMDC) such as MoS$_2$. Staggered potential only shift the energy 
of $n=0$ LL to $-\eta m$ without changing energies of other LLs. Also, it does not change the 
wave functions of LLs.
For example, when the Fermi energy lies between $n=0$ LLs and $n=1$ LLs 
(both at K and K' points) as illustrated in Fig.~\ref{fig: LLs}(a), nonzero shift current flows by tuning the photon frequency to either 
$\hbar \omega_c - m$ (resonant at K valley) or $\hbar \omega_c+ m$ (resonant at K') valley.

Next, we move on to shift current in LLs on the surface of topological insulators(TIs).
The Hamiltonian is written as
\cite{Fu09}
\begin{align}
H=v_F (p_x \sigma_y  - p_y \sigma_x ) + \frac{\lambda}{2}(p_+^3 + p_-^3) \sigma_z,
\end{align}
where $v_F$ is the Fermi velocity, $\lambda$ is the strength of trigonal warping effect, and 
$p_\pm= p_x \pm i p_y$.
The Hamiltonian has $C_3$ rotation symmetry and reflection symmetry along the $x$ direction 
[$(k_x,k_y) \to (-k_x, k_y)$] with $R_x= i\sigma_x$. With $C_{3v}$ symmetry, 
nonzero components of second order nonlinear conductivity are 
$\sigma_{yyy}=-\sigma_{yxx}$.
The Hamiltonian in the magnetic field is again obtained by replacing the momentum operators with creation/annihilation operators (for details, see Appendix \ref{app: LL}).
The nonlinear conductivity $\sigma_{yyy}$ in the lowest order in the trigonal warping is given by
\begin{align}
\sigma_{yyy}(\omega)&= - \frac{3 e^3}{\sqrt{2} \omega^2 \ell^3} v_F \lambda
\delta(\omega-\omega_{10}),
\end{align}
when the Fermi energy is located between $n=0$ and $n=1$ LLs, and the photon frequency is close to the cyclotron energy ($\omega \sim \omega_c$). \footnote{We note that the nonlinear conductivity in TI surface states show a different $B$ dependence ($\propto B^{\frac 3 2}$ from $\ell \propto 1/\sqrt{B}$) compared to graphene ($\propto B$). This is a consequence of different forms of trigonal warping ($\propto p^3$ in TI and $\propto p^2$ in graphene).}
A general expression for $\sigma_{yyy}$ can be found in Appendix \ref{app: LL}, and
it shows that the two interband transitions $-n \to n+1$ and $-n-1 \to n$ (that have the same resonance
frequency) make opposite contributions to shift current (see Fig.~\ref{fig: LLs}(b)). In Fig.~\ref{fig: LLs}(c), optical absorption shows 
consecutive peaks from interband contributions while shift current shows only one peak at the 
resonance $0 \to 1$.

We can estimate photocurrent supported by the LLs as follows. 
For $B=1\t{ T}$, the cyclotron energy is given by $\hbar \omega= 38 \t{meV}$ (graphene \cite{Ando98,Koshino-McCann09,McCann-Koshino13}), $22 \t{meV}$ (MoS$_2$ \cite{Kormanyos13}) and $\hbar \omega=14 \t{meV}$ (TI \cite{Fu09}) using material parameters from the references, 
which falls into terahertz regime.
For typical relaxation time of $\tau=1 \t{ ps}$, the 2D nonlinear conductivity are estimated as
 $\sigma= 7.3 \times 10^{-12} \t{ A m/V}^2$ (graphene),
$ 5.5 \times 10^{-12} \t{ A m/V}^2$ (MoS$_2$),
and $ 1.7 \times 10^{-11} \t{ A m/V}^2$ (TI).
These values exceed $\sigma$ for typical 2D shift current material monolayer GeS $\sigma \sim 10^{-14} \t{ A m/V}^2$ \cite{Rangel17}.
We can estimate photocurrent $J$ generated by light intensity $I$ from the formula $J=\kappa I$ with $\kappa = 2 \sigma/c \epsilon_0$  \cite{Cook17}.
This allows us to convert 
$\sigma= 1.0 \times 10^{-11} \t{ A m/V}^2$ to
$\kappa = 7.6 \times 10^{-5} \t{ (A/m)} / (\t{W/cm}^2)$.
For example, when TI of sample size 1 mm is irradiated with $I=1 \t{ W/cm}^2$ under $B=1 \t{ T}$, the shift current from LLs reaches 100 nA, which is much larger than photocurrent of 10 - 100 pA that has been observed for a TI thin film \cite{Okada16}

\begin{figure}
\includegraphics[width=0.8\linewidth]{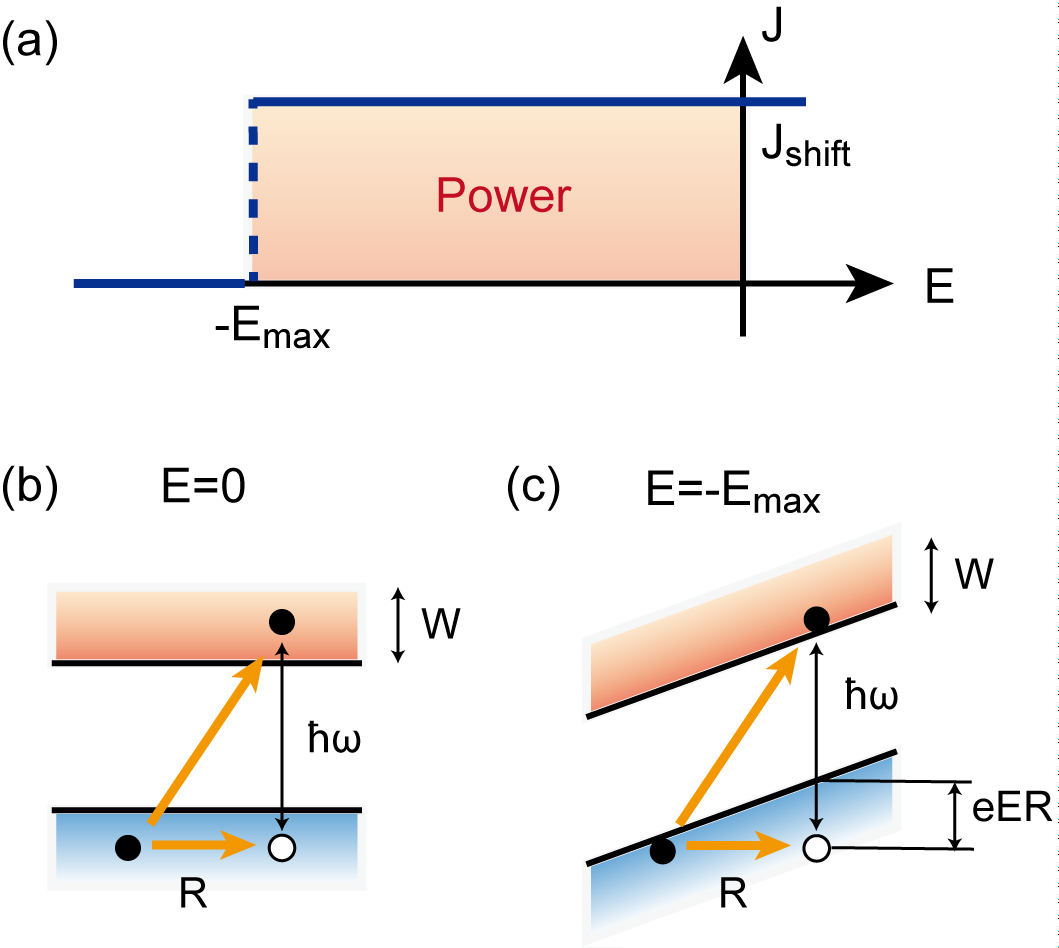}
\caption{\label{fig: efficiency}
(a) $I-V$ characteristic of flat band systems.
(b) Photoexcitation in zero electric field.
(c) Photoexcitation in $E=-E_\t{max}$. Real space shift $R$ causes energy shift of $eER$. 
}
\end{figure}

Finally, we discuss the efficiency of shift current of the LLs.
While $\sigma_E$ becomes zero due to the flat band nature of the LLs, there exists a maximum electric field $E_\t{max}$ that we can apply to LL photovoltaics as illustrated in Fig.~\ref{fig: efficiency}.
In the electric field $E_\t{dc}$, the real space shift of an electron quantified by shift vector $R$ causes energy shift of $eE_\t{dc}R$. In order that the photoexcitation still takes place in the presence of $E_\t{dc}$, this energy shift $eE_\t{dc}R$ should not exceed the band width $W$ of valence and conduction bands (see Fig.~\ref{fig: efficiency}(c)).
Therefore, the maximum electric field is determined by $E_\t{max}=W/eR$.
In the case of LLs, the band width is given by $W\simeq \hbar/\tau$ with relaxation time $\tau$.
(Here we assume that the sample is irradiated with monochromatic light of cyclotron frequency and the LLs have uniform level broadening of $W$, where optical transition equally takes place as far as $|E_\t{dc}|<E_\t{max}$.)
Now we estimate monochromatic power conversion efficiency $r=P_\t{out}/P_\t{in}$ for monochromatic light with the cyclotron frequency.
Here, $P_\t{out}$ is the power generated by shift current photovoltaics that is given by $P_\t{out}=J_\t{shift} E_\t{max}$,
and $P_\t{in}$ is the power absorbed by the sample that is given by $P_\t{in}=\sigma^{(1)}(\omega) |E(\omega)|^2$ where $\sigma^{(1)}(\omega)$ is the linear conductivity quantifying the absorption rate of the monochromatic light.
By using a rough approximation for Eq.~(\ref{eq: J shift}) as $J_\t{shift} \simeq (e/\hbar \omega)|E(\omega)|^2 \sigma^{(1)} R$,
we obtain the conversion efficiency for flat band systems as
\begin{align}
r \simeq \frac{W}{\hbar \omega} \simeq \frac{1}{\omega \tau}.
\end{align}
For $B=1\t{T}$,
 the efficiency $r$ becomes as high as $r=11\%$ (graphene), $18\%$ (MoS$_2$), and $29 \%$ (TI) with a typical level broadening $W=4 \t{meV}$ from impurity scattering.

\textit{Discussions ---}
We have studied $I-V$ characteristics and shot noise in shift current photovoltaics. 
The derived formulae for them indicate that the slope in the $I-V$ characteristic and the shot noise are strongly disorder and temperature dependent through their $\tau$ dependence,
while zero bias shift current is independent of disorder and temperature.
The formulae also show that it is possible to suppress both the slope in $I-V$ characteristics and the nonequilibrium noise simultaneously 
by reducing the band widths. 

Flat band systems are the ideal laboratory to realize this situation.
We have proposed that the LLs of graphene and surface states of 3D TIs are promising candidates for this purpose.
We predict that the shift current of LLs is observable with essentially no current fluctuation and with high monochromatic power conversion efficiency. This will offer the sharpest experimental test of the geometrical nature of the shift current. 
For application as photodetectors for weak intensity light, we also need to consider noise that comes from quantum statistics of photons, which is left for future studies.

\begin{acknowledgments}
We thank J. Orenstein, and J.E. Moore for fruitful discussions.
TM was supported by the Gordon and Betty Moore Foundation's EPiQS Initiative Theory Center Grant to UC Berkeley, and the Quantum Materials program at LBNL funded by the US Department of Energy under Contract No. DE-AC02-05CH11231. 
MN was supported by PRESTO, JST (No. JPMJPR16R5). 
MK was supported by CREST, JST (No. JPMJCR16F1).
NN was supported by JSPS KAKENHI (No. 18H03676), and ImPACT Program of Council for Science, Technology and Innovation (Cabinet office, Government of Japan, 888176).
\end{acknowledgments}

\appendix
\input{SI-condmat.tex}

\bibliography{noise}

\end{document}

%% file: SI-condmat.tex
\section{Floquet two band model \label{app: floquet}}

\begin{figure}[tb]
\includegraphics[width=0.5\linewidth]{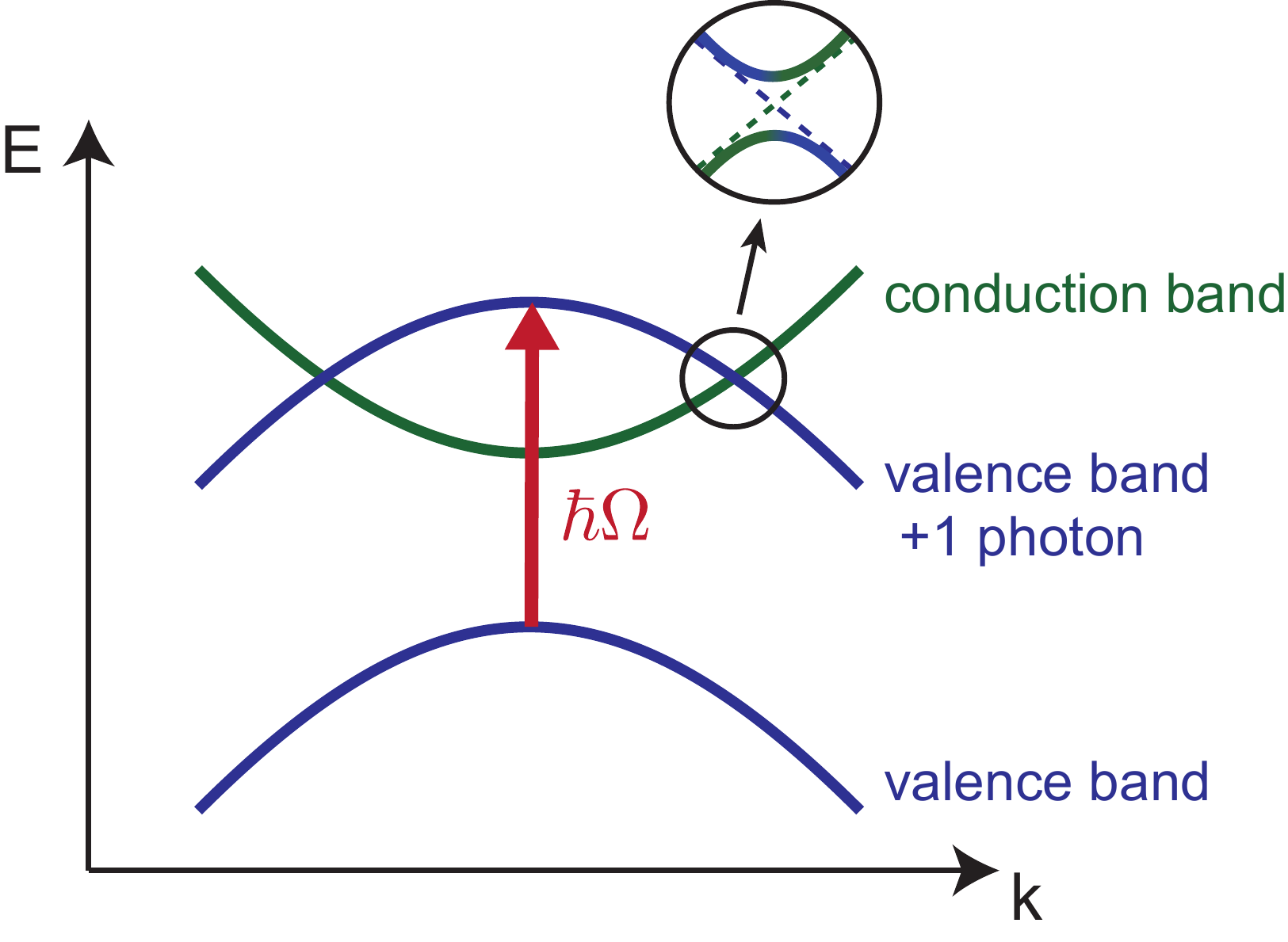}
\caption{\label{fig: floquet anticrossing}
Schematic picture of Floquet two band model. We consider anticrossing of the valence band with one photon and the conduction band with zero photon. [From Ref.~\cite{Morimoto-Nagaosa16}]
}
\end{figure}

Floquet two band model can describe shift current in a concise way as we briefly explain below \cite{Morimoto-Nagaosa16,Nagaosa-Morimoto17}.
We focus on the valence band with one photon and the conduction band with zero photon as illustrated in Fig.~\ref{fig: floquet anticrossing}.
The Floquet Hamiltonian is given by
\begin{align}
H_F&=
\bm d \cdot \bm \sigma,
\end{align}
with
\begin{align}
d_x - id_y &= A v_{12}, \\
d_z &= \frac{\epsilon_1-\epsilon_2 + \omega}{2},
\label{eq: def d}
\end{align}
with $A=E/\omega$.
Here we set $\hbar=1,e=1$ for simplicity.

In the Keldysh Green's function formalism,
the Green's function $G$ has a matrix form that is given by the Dyson equation as
\begin{align}
\begin{pmatrix}
G^R & G^K \\
0 & G^A
\end{pmatrix}^{-1}
&=
\omega-H -
\begin{pmatrix}
\Sigma^R & \Sigma^K \\
0 & \Sigma^A
\end{pmatrix}.
\label{eq: Dyson}
\end{align}
We introduce the self energy $\Sigma$ through coupling to heat bath (with wide spectrum and constant DOS, for simplicity), which stabilize nonequilibrium electron distribution in the Floquet two band model. In this setup, the self energy $\Sigma$ is given by \cite{Rammer86,Jauho94,Aoki-RMP14}
\begin{align}
\begin{pmatrix}
\Sigma^R & \Sigma^< \\
0 & \Sigma^A
\end{pmatrix}
 &=
i\Gamma
\begin{pmatrix}
-1/2 & f \\
0 & 1/2
\end{pmatrix}.
\end{align}
Here $f$ is the Fermi distribution function, for which we assume zero temperature form $f(\omega)=\theta(E_F-\omega)$ ($\theta$: step function, $E_F$: Fermi energy).
By using the Dyson equation,
Retarded and advanced Green's functions are given by
\begin{align}
G^{R/A}(\omega) &= \omega - H_F \pm i \frac{\Gamma}{2}. 
\end{align}
The lesser and greater Green's functions are given by \cite{Jauho94},
\begin{align}
G^<&= G^R \Sigma^< G^A, \\
G^>&= G^R \Sigma^> G^A, 
\label{eq: G < and G>}
\end{align}
with
\begin{align}
\Sigma^<&= i \Gamma \frac{1+\sigma_z}{2}, \\
\Sigma^>&= - i \Gamma \frac{1-\sigma_z}{2}.
\end{align}

The velocity operator in the Floquet two band model is obtained by taking $k$ derivative of the Floquet Hamiltonian as
\begin{align}
v_k&=\frac{d H_F}{dk} \equiv \bm b \cdot \bm \sigma,
\end{align}
with
\begin{align}
b_x - ib_y &= A (\partial_k v)_{12}, \\
b_z &= \frac{v_{11}- v_{22}}{2}.
\label{eq: def b}
\end{align}
Shift current $J_\t{shift}$ can be obtained as
\begin{align}
J_\t{shift} 
&= -i \int \frac{d\omega}{2\pi} \int [dk] \t{tr}[v_k G^<_0] \n
&= \int[dk] \frac{2 (- d_x b_y + d_y b_x) \Gamma}{4 d_z^2 + \Gamma^2},
\end{align}
where $[dk] \equiv dk/(2\pi)^d$ with the dimension $d$.
By using Eq.~(\ref{eq: def d}) and Eq.~(\ref{eq: def b}), and restoring $e$ and $\hbar$,
we obtain nonlinear conductivity $\sigma^{(2)}(\omega)$ that is defined by $J_\t{shift}= \sigma^{(2)}(\omega) |E(\omega)|^2$ as
\begin{align}
\sigma^{(2)}(\omega)
&= 
\frac{2\pi e^3}{\hbar^2 \omega^2}
\int[dk] \t{Im} \left[ \left(\frac{\partial v}{\partial k} \right)_{12} v_{21} \right] \delta(\omega_{21}-\omega),
\end{align}
where $\omega_{21}=(\epsilon_2-\epsilon_1)/\hbar$.
Here we used $\Gamma/(4d_z^2+\Gamma^2)= \delta(\epsilon_2-\epsilon_1-\hbar\omega)$.
This can be rewritten as \cite{Morimoto-Nagaosa16}
\begin{align}
\sigma^{(2)}(\omega)
&= 
\frac{2\pi e^3}{\hbar^2 \omega^2}
\int[dk] |v_{12}|^2 R \delta(\omega_{21}-\omega),
\end{align}
with
the shift vector $R$ given by
\begin{align}
R= \t{Im}\left[\frac{(\partial_k v)_{12}}{v_{12}}\right] 
= \t{Im}[\partial_k (\log v_{12})] + a_1 - a_2,
\end{align}
where $a_{i}$ is the Berry connection for the band $i$.
We note that the unit of nonlinear conductivity in 3D is 
$[\sigma^{(2)}] = [\t{A/V}^2]$ in SI unit.

\section{Derivation of I-V characteristics of shift current photovoltaics \label{app: I-V}}
In this section, we derive $I-V$ characteristics of shift current photovoltaics by combining gauge invariant formalism of Keldysh Green's function and Floquet two band model.

\subsection{Gauge invariant formalism of Keldysh Green's function}
We study a system in an external electric field by using Keldysh Green's function 
and gradient 
expansion.
In the presence of an external electric field $E$, the Green's function and the self energy are 
expanded with respect to $E$ as \cite{Onoda06,Sugimoto08,Morimoto18}
\begin{align}
G(\pi) &= G_0(\pi) + \frac{E}{2} G_E(\pi) +O(E^2), \\
\Sigma(\pi) &= \Sigma_0(\pi) + \frac{E}{2} \Sigma_E(\pi) +O(E^2),
\end{align}
where we set $\hbar=1,e=1$ for simplicity.
The unperturbed part of the Green's function $G_0$ is given by
Eq.~(\ref{eq: Dyson}),
and the linear order correction to the Green's function $G_E$ is given by
\begin{widetext}
\begin{align}
G_E=G_0 \left[
\Sigma_E + \frac{i}{2} 
\left( 
(\partial_{\omega} G_0^{-1}) G_0  (\partial_{k} G_0^{-1})
-
(\partial_{k} G_0^{-1}) G_0 (\partial_{\omega} G_0^{-1})
\right) \right] G_0.
\label{eq:GE}
\end{align}
\end{widetext}

\subsection{Application to Floquet two band model}
We apply the gauge invariant formalism to Floquet two band model and compute $\sigma_E$ by expanding current response with dc electric field $E$ as
\begin{align}
J= J_\t{shift} + \sigma_E E. 
\end{align}
When we assume that the bath is large and its electronic structure is not modified by $E$, we can 
write $G^<_E$ as
\begin{align}
G^<_E &= - \frac i 2 G_0[ G_0 v_k - v_k G_0 ] G_0,
\end{align}
where the $G_0$ has a matrix structure given by Eq.~(\ref{eq: Dyson}).
We note that the contribution involving 
$\partial_\omega \Sigma^<_0 = i \Gamma \partial_\omega f$ does not appear, because there 
is no Fermi surface in the original band structure (we assumed an insulator).

The linear correction to current expectation value $\sigma_E$ is given by
\begin{align}
\sigma_E &= -i \int \frac{d\omega}{2\pi} \int [dk] \t{tr}[v_k G^<_E] \n
&=
\int [dk] \frac{4 \pi b_z (b_x d_x + b_y d_y)}{\Gamma^2} \delta(d_z)  +O(A^3)
.
\end{align}
By restoring $e$ and $\hbar$, this expression can be rewritten as
\begin{align}
\sigma_E&= \frac{4 \pi e^4}{\hbar^3 \omega^2} |E(\omega)|^2 \tau^2 \int [dk]
|v_{12}|^2 (v_{11}-v_{22}) R' \delta(\omega_{21}-\omega),
\end{align}
with
$R'=\t{Re}[(\partial_k v)_{12}/v_{12}]$.
$R'$ has a dimension of length and is a similar quantity to shift vector $R$.
We note that the unit of $\sigma_E$ in 3D is 
$[\sigma_E] = [\t{A/Vm}]$ in SI unit.

\section{Calculation of current noise in the bulk \label{app: noise}}
We compute current noise in the bulk. We consider autocorrelation of current operator defined 
in one point $(x=0)$ in the bulk crystal. The local current operator is given by \cite{Marel15}
\begin{align}
v_{loc}=\frac 1 L \sum_{k,k'} \frac{v_k + v_{k'}}{2}c_k^\dagger c_{k'},
\end{align}
where $L$ is the sample size. We note that $k$ and $k'$ should satisfy $|k-k'|<1/l$ with the size of the electrode $l$ since the $k$ transfer at the electrode is bounded by the size of the electrode due to uncertainty principle.
The current noise is given by zero frequency component of the autocorrelation function of $v_{loc}$, and is written as
\begin{widetext}
\begin{align}
S &=\int dt (\braket{v_{loc}(t) v_{loc}(0)} - \braket{v_{loc}}^2) \n
&=\frac{1}{4 L^2} \int dt \sum_{k,k', k'', k'''} (v_k + v_{k'})(v_{k''} + v_{k'''}) 
(
\braket{c_{k'''}^\dagger(t) c_{k''}(t) c_{k'}^\dagger(0) c_{k}(0)}
-
\braket{c_{k'''}^\dagger c_{k''}} \braket{c_{k'}^\dagger c_{k}}
).
\end{align}
In the case of noninteracting electrons, we obtain
\begin{align}
S&=\frac{1}{4L^2} \int dt \sum_{k,k', k'', k'''} (v_k + v_{k'})(v_{k''} + v_{k'''})
\braket{c_{k'''}^\dagger(t) c_{k}(0)} \braket{c_{k''}(t) c_{k'}^\dagger(0)} \n
&=\frac{1}{4L^2} \int \frac{d\omega}{2\pi} \sum_{k,k', k'', k'''} (v_k + v_{k'})(v_{k''} + v_{k'''})
G^<(\omega,k) G^>(\omega,k') \delta_{k''', k} \delta_{k'',k'} \n
&=\frac{1}{4} \int \frac{d\omega}{2\pi} [dk] [dk'] (v_k + v_{k'})^2
G^<(\omega,k) G^>(\omega,k').
\label{eq: shot noise}
\end{align}
\end{widetext}

Now we study shot noise in shift current photovoltaics by using Floquet two band model.
Equation~(\ref{eq: shot noise}) with Green's functions in Eq.~(\ref{eq: G < and G>}) leads to
\begin{align}
S&= S_1 + S_2 + S_3 + O(A^3),
\end{align}
with
\begin{align}
S_1&= \frac{1}{\Gamma} \int [dk] b_z^2 (d_x^2+ d_y^2) \delta(d_z) \int dk' \delta(d_z'), \\
S_2&= \int [dk] b_z (d_x b_y - d_y b_x) \delta(d_z) \int dk' \delta(d_z'), \\
S_3&= \frac{1}{4} \int [dk] [dk'] [(b_x+b_x')^2+ (b_y+b_y')^2] \delta(d_z + d_z'),
\end{align}
where $\bm{d'}=\bm d(k')$ and $\bm{b'}=\bm b(k')$.
Here we only kept contributions from optical absorption that involve delta functions such as $\delta(d_z)$.
Under time reversal symmetry $\mathcal{T}=K$ ($K$: complex conjugation),  $\bm{d}$ and $\bm b$ transform as
\begin{align}
(d_x(k), d_y(k), d_z(k)) &\to (d_x(-k), -d_y(-k), d_z(-k)), \\
(b_x(k), b_y(k), b_z(k)) &\to (-b_x(-k), b_y(-k), -b_z(-k)).
\end{align}
We find from these that $S_1$ and $S_3$ are even under the TRS, while $S_2$ is odd and vanishes after 
integration over $k$.
The integrand of $S_3$ is nonzero only when $d_z(k)=-d_z(k')$. The momentum $k$ satisfies this 
condition only for $|k|< l$, which means $S_3$ is of the order of $l$ and negligible for a large enough 
electrode.
Thus only $S_1$ contributes to the noise.

The noise $S$ can be rewritten with Eq.~(\ref{eq: def d}) and Eq.~(\ref{eq: def b}) as
\begin{align}
S=S_1= \frac{E^2}{\omega^2} \tau \int [dk] |v_{11}-v_{22}| |v_{12}|^2 \delta(\omega_{21} - \omega),
\end{align}
by noticing $\delta(d_z)=\delta(k)/|b_z|$.
Restoring $e$ and $\hbar$ leads to
\begin{align}
S= \frac{e^4}{\hbar^2 \omega^2} E^2 \tau \int [dk]  |v_{11}-v_{22}| |v_{12}|^2 \delta(\omega_{21} - \omega).
\label{eq: S shift}
\end{align}
We note that the unit of $S$ is [C$^2$/s] in 1D.

\section{Shift current in Landau levels \label{app: LL}}

\subsection{Landau levels in graphene}
We study shift current and shot noise in graphene Landau levels (LLs).
Since LLs have flat band dispersion ($v_{11}=v_{22}=0$), we expect that the noise is suppressed in LLs.
In order to take into account the effect of inversion breaking in graphene,
we consider Dirac Hamiltonian with a trigonal warping term \cite{Ando98,Koshino-McCann09,McCann-Koshino13},
\begin{align}
H &= v_F
\begin{pmatrix}
0 & \pi^\dagger \\
\pi & 0 \\
\end{pmatrix}
+ \eta \kappa
\begin{pmatrix}
0 & \pi^2 \\
(\pi^\dagger)^2 & 0 \\
\end{pmatrix},
\end{align}
with Fermi velocity $v_F$, strength of trigonal warping $\lambda$,
$\pi=\hbar(\eta k_x + i k_y)$, $\pi^\dagger=\hbar(\eta k_x - i k_y)$, and $\eta=\pm 1$ for K and K' valleys, respectively.
In the presence of magnetic field,
the momentum operators are written as
$\pi^\dagger=(\sqrt 2 \hbar/\ell)a^\dagger$ at K and 
$\pi^\dagger=(\sqrt 2 \hbar/\ell)a$ and K' 
with raising and lowering operators $a^\dagger$ and $a$, and the magnetic length $\ell=\sqrt{\hbar/eB}$. 
In the valley K with $\kappa=0$,
the wave functions of LLs are given as
\begin{align}
H \psi_{n} &= \t{sgn}(n) \sqrt{|n|} \hbar \omega_c, \\
\psi_n &= C_n
\begin{pmatrix}
\phi_{|n|} \\
\t{sgn}(n) \phi_{|n|-1}
\end{pmatrix},
\end{align}
with $s =\pm 1$ and $C_n=1/\sqrt{2} (n \neq 0), 1 (n=0)$.
Here, $\omega_c=v_F\sqrt{2eB/\hbar}$ is the cyclotron frequency, and $\phi_n$ is the $n$th Landau wave function (with $\phi_{-1}\equiv 0$).
For the K' valley, the wave function is obtained by exchanging the two components.
In the first order in the perturbation with $\kappa$, the wave function is modified as 
$\psi_n \to \psi_n+\delta \psi_n$ with
\begin{align}
\delta \psi_n &=
\kappa(c_1 \psi_{n+3} + c_2 \psi_{n-3}),
\end{align}
where $c_1$ and $c_2$ are constants of the order of 1/$\hbar \omega_c$. $\delta \psi_n$ gives higher order corrections in $\kappa$ to nonlinear conductivity, and we can neglect it as long as we focus on the leading order term in $\kappa$.

The paramagnetic current operator in the $x$ direction is given by
\begin{align}
v=\frac{1}{\hbar}\frac{\partial H}{\partial k_x}=
v_F
\begin{pmatrix}
0 & 1 \\
1 & 0 \\
\end{pmatrix}
+ 2 \eta \kappa
\begin{pmatrix}
0 & \pi \\
\pi^\dagger & 0 \\
\end{pmatrix}.
\end{align}
The diamagnetic current operator in the $y$ direction with $E_x$ is given by
\begin{align}
\partial_{k_y}v &=
-2 \eta \hbar \kappa 
\begin{pmatrix}
0 & -i \\
i & 0 \\
\end{pmatrix}.
\end{align}

Now let us consider shift current supported by the graphene LLs. We suppose that the Fermi 
level is located between $n=0$ and $n=1$ LLs, and the sample is irradiated with linearly 
polarized light along the $x$ direction with the photon energy close to $\hbar \omega_c$.
In this case, the photoexcitation takes place from $n=0$ LL to $n=1$ LL. The shift current 
response $J_y = \sigma_{yxx}(\omega) |E_x(\omega)|^2$ is given by
\begin{align}
\sigma_{yxx}(\omega)=\frac{2 \pi e^3}{\hbar^2 \omega^2} \t{Im}[\bra{\psi_0} \partial_{k_y} v_x \ket{\psi_1} \bra{\psi_1} v_x \ket{\psi_0}] 
D(\omega).
\end{align}
Here, $D(\omega)$ is the joint density of states given by 
$D(\omega)=\frac{1}{2 \pi \ell^2}\delta(\omega-\omega_{10})$, 
where we replace the delta function with a Lorentzian as $\delta(\omega-\omega_{10}) \to (1/\pi)\frac{1/\tau}{(\omega-\omega_c^2)+1/\tau^2}$ in numerical calculations.
Since the current matrix elements are given by
\begin{align}
\bra{\psi_1} v_x \ket{\psi_0} &= \frac{v_F}{\sqrt 2}, \\
\bra{\psi_0} \partial_{k_y} v_x \ket{\psi_1} &= i \eta \sqrt{2} \kappa,
\end{align}
the shift current is given by
\begin{align}
\sigma_{yxx}(\omega)= \eta \frac{e^3}{\hbar \omega^2 \ell^2} v_F \kappa \delta(\omega-\omega_{10}).
\label{eq: sigma graphene}
\end{align}
We note that we obtain the same shift current for the transition $-1 \to 0$.
Since $\sigma_{yxx}$ is proportional to $\eta$, the shift current response vanishes when the optical transition $0 \to 1$ 
takes place both at K and K' valleys.
This is natural since the original Hamiltonian has inversion symmetry 
$\mathcal{I}=\sigma_x \tau_x$ (where $\sigma$ and $\tau$ are Pauli matrices acting on 
sublattice and valley degrees of freedom), and
nonzero shift current requires breaking of inversion symmetry. Staggered potential $m \sigma_z \tau_0$ breaks inversion symmetry. Introducing staggered potential only changes the energy 
of $n=0$ LL to $-\eta m$ while the same expression for shift current [Eq.~(\ref{eq: sigma graphene})] holds.
In this setup, the two optical transitions at $\omega=\omega_c \pm m /\hbar$ (resonant either at K and K' valleys) support nonzero shift current (with opposite signs with each other).

\subsection{Landau levels in TI surface}
We study shift current in LLs on the surface of topological insulators(TIs).
The Hamiltonian is written as
\cite{Fu09}
\begin{align}
H=v_F (p_x \sigma_y  - p_y \sigma_x ) + \frac{\lambda}{2}(p_+^3 + p_-^3) \sigma_z,
\end{align}
where $v_F$ is the Fermi velocity, $\lambda$ is the strength of trigonal warping effect, and 
$p_\pm= p_x \pm i p_y$.
The Hamiltonian has $C_3$ rotation symmetry and reflection symmetry along the $x$ direction 
($(k_x,k_y) \to (-k_x, k_y)$) with $R_x= i\sigma_x$. With $C_{3v}$ symmetry, 
nonzero components of second order nonlinear conductivity are 
$\sigma_{yyy}=-\sigma_{yxx}$ \cite{Sturman}.
In the magnetic field, the momentum operator is written as 
$p_+ = \pi = (\sqrt{2}\hbar/\ell) a$.
The Hamiltonian is rewritten with creation/annihilation operators as
\begin{align}
H=
\begin{pmatrix}
\frac{\lambda}{2}(\pi^3 + (\pi^\dagger)^3) & -i \pi^\dagger \\
i \pi & -\frac{\lambda}{2}(\pi^3 + (\pi^\dagger)^3)
\end{pmatrix}.
\end{align}
When we set $\lambda=0$, the wave function is given by
\begin{align}
E_n&=\t{sgn}(n) \sqrt{|n|} \hbar \omega_c, \\
\psi_n&=C_n
\begin{pmatrix}
\phi_{|n|} \\
i ~ \t{sgn}(n) \phi_{|n|-1}
\end{pmatrix},
\end{align}
with $C_n=1/\sqrt{2} (n \neq 0), 1 (n=0)$.
The wave function has a correction of the order of $\lambda$ that we can neglect since we focus on the leading order contribution in $\lambda$.

Now we study the nonlinear conductivity $\sigma_{yyy}$.
Paramagnetic current and diamagnetic current operators are given by
\begin{align}
v_{y} &= - v_F \sigma_x + \frac{3 \lambda}{2}(i p_+^2 -i p_-^2)\sigma_z, \\
\partial_{k_y}v_{yy} &= - 3 \hbar \lambda (p_+ + p_-) \sigma_z, 
\end{align}
Since the shift current vanishes for $\lambda=0$ due to the rotation symmetry $C_{\infinity}$,
the leading order contribution to $\sigma_{yyy}$ is $ O(\lambda)$.
Since the diamagnetic current is already proportional 
to $\lambda$, we can use the wave function for 
$\lambda=0$ to evaluate matrix elements.
If we focus on the optical transition between $n=0$ and $n=1$ LLs,
the current response is given by
\begin{align}
\sigma_{yyy}(\omega)= \frac{2 \pi e^3}{\hbar^2 \omega^2} \t{Im}[\bra{\psi_0} \partial_{k_y} v_{y} \ket{\psi_1} \bra{\psi_1} 
v_{y} \ket{\psi_0}] D(\omega),
\end{align}
with $D(\omega)=\frac{1}{2\pi \ell^2}\delta(\omega-\omega_{10})$.
By using
\begin{align}
\bra{\psi_1} v_{y} \ket{\psi_0} 
&=
i \frac{v_F}{\sqrt{2}} + O(\lambda), \\
\bra{\psi_0} \partial_{k_y} v_{y} \ket{\psi_1}
&=-\frac{3 \hbar^2 \lambda}{\ell} + O(\lambda^2),
\end{align}
we obtain
\begin{align}
\sigma_{yyy}(\omega)&= - \frac{3 e^3}{\sqrt{2} \omega^2 \ell^3} v_F \lambda
\delta(\omega-\omega_{10}).
\end{align}
This expression holds when $\omega \sim \omega_c$ and the Fermi energy is located between $n=0$ and $n=1$ LLs.

Below we give a complete calculation for other LLs.
The general matrix elements are given by
\begin{align}
\bra{\psi_{s'(n+1)}} v_{y} \ket{\psi_{sn}} 
&=
i s' C_n C_{n+1} v_F + O(\lambda), 
\end{align}
and
\begin{align}
&\bra{\psi_{sn}}  \partial_{k_y} v_{y} \ket{\psi_{s'(n+1)}} \n
&=-\frac{3 \sqrt{2} C_n C_{n+1} \hbar^2 \lambda}{\ell} (s s' \sqrt{n} + \sqrt{n+1}) + O(\lambda^2),
\end{align}
where $s,s'=\pm 1$ and we set $n \ge 0$.
The nonlinear conductivity $\sigma_{yyy}$ is given by a sum of contributions from the optical 
responses $\psi_{sn} \to \psi_{s'(n+1)}$ as
\begin{widetext}
\begin{align}
\sigma_{yyy}(\omega)&= 
- \frac{3 \sqrt{2} e^3}{\omega^2 \ell^3} v_F \lambda
\sum_{s,s',n}
C_n^2 C_{n+1}^2
s'(s s' \sqrt{n}+\sqrt{n+1}) \n
&\qquad \qquad \qquad \qquad 
\times [f(E_{sn})(1-f(E_{s'(n+1)})) \delta(\omega-(s'\sqrt{n+1}-s\sqrt{n})\omega_c)  \n
&\qquad \qquad \qquad \qquad \quad
- f(E_{s'(n+1)})(1-f(E_{sn}))\delta(\omega-(s\sqrt{n}-s'\sqrt{n+1})\omega_c)],
\end{align}
\end{widetext}
where $f(E)$ is the Fermi distribution function that ensures optical transitions between occupied and unoccupied states.
We notice that two interband transitions $-n \to n+1$ and $-n-1 \to n$ have the same resonant 
frequency but opposite contributions to shift current. 
For comparison, the linear conductivity $\sigma_{yy}(\omega)$ in quantum Hall system is generally given by
\begin{align}
\sigma_{yy}(\omega)= \frac{2 \pi e^2}{\hbar \omega} \sum_{m,n}|\bra{\psi_m} v_{y} \ket{\psi_n} |^2 D_{mn}(\omega),
\end{align}
with $D_{mn}(\omega)=(1/2\pi\ell^2) \delta(\omega-\omega_{mn})$,
and $\sigma_{yy}(\omega)$ for TI LLs is written as
\begin{widetext}
\begin{align}
\sigma_{yy}(\omega)&=  \frac{e^2 v_F^2}{\hbar \omega \ell^2} 
\sum_{s,s',n}
C_n^2 C_{n+1}^2 
[f(E_{sn})(1-f(E_{s'(n+1)})) \delta(\omega-(s'\sqrt{n+1}-s\sqrt{n})\omega_c)  \n
& \qquad \qquad \qquad \qquad \qquad
+ f(E_{s'(n+1)})(1-f(E_{sn}))\delta(\omega-(s\sqrt{n}-s'\sqrt{n+1})\omega_c)].
\end{align}
\end{widetext}
We note that the transitions $-n \to n+1$ and $-n-1 \to n$ add up in the linear conductivity.



%% file: condmat1130.bbl
\begin{thebibliography}{33}%
\makeatletter
\providecommand \@ifxundefined [1]{%
 \@ifx{#1\undefined}
}%
\providecommand \@ifnum [1]{%
 \ifnum #1\expandafter \@firstoftwo
 \else \expandafter \@secondoftwo
 \fi
}%
\providecommand \@ifx [1]{%
 \ifx #1\expandafter \@firstoftwo
 \else \expandafter \@secondoftwo
 \fi
}%
\providecommand \natexlab [1]{#1}%
\providecommand \enquote  [1]{``#1''}%
\providecommand \bibnamefont  [1]{#1}%
\providecommand \bibfnamefont [1]{#1}%
\providecommand \citenamefont [1]{#1}%
\providecommand \href@noop [0]{\@secondoftwo}%
\providecommand \href [0]{\begingroup \@sanitize@url \@href}%
\providecommand \@href[1]{\@@startlink{#1}\@@href}%
\providecommand \@@href[1]{\endgroup#1\@@endlink}%
\providecommand \@sanitize@url [0]{\catcode `\\12\catcode `\$12\catcode
  `\&12\catcode `\#12\catcode `\^12\catcode `\_12\catcode `\%12\relax}%
\providecommand \@@startlink[1]{}%
\providecommand \@@endlink[0]{}%
\providecommand \url  [0]{\begingroup\@sanitize@url \@url }%
\providecommand \@url [1]{\endgroup\@href {#1}{\urlprefix }}%
\providecommand \urlprefix  [0]{URL }%
\providecommand \Eprint [0]{\href }%
\providecommand \doibase [0]{http://dx.doi.org/}%
\providecommand \selectlanguage [0]{\@gobble}%
\providecommand \bibinfo  [0]{\@secondoftwo}%
\providecommand \bibfield  [0]{\@secondoftwo}%
\providecommand \translation [1]{[#1]}%
\providecommand \BibitemOpen [0]{}%
\providecommand \bibitemStop [0]{}%
\providecommand \bibitemNoStop [0]{.\EOS\space}%
\providecommand \EOS [0]{\spacefactor3000\relax}%
\providecommand \BibitemShut  [1]{\csname bibitem#1\endcsname}%
\let\auto@bib@innerbib\@empty
\bibitem [{\citenamefont {Bloembergen}(1996)}]{Bloembergen}%
  \BibitemOpen
  \bibfield  {author} {\bibinfo {author} {\bibfnamefont {N.}~\bibnamefont
  {Bloembergen}},\ }\href@noop {} {\emph {\bibinfo {title} {Nonlinear
  optics}}}\ (\bibinfo  {publisher} {World Scientific, Singapore},\ \bibinfo
  {year} {1996})\BibitemShut {NoStop}%
\bibitem [{\citenamefont {Boyd}(2003)}]{Boyd}%
  \BibitemOpen
  \bibfield  {author} {\bibinfo {author} {\bibfnamefont {R.~W.}\ \bibnamefont
  {Boyd}},\ }\href@noop {} {\emph {\bibinfo {title} {Nonlinear optics}}}\
  (\bibinfo  {publisher} {Academic press, London},\ \bibinfo {year}
  {2003})\BibitemShut {NoStop}%
\bibitem [{\citenamefont {von Baltz}\ and\ \citenamefont
  {Kraut}(1981)}]{Kraut}%
  \BibitemOpen
  \bibfield  {author} {\bibinfo {author} {\bibfnamefont {R.}~\bibnamefont {von
  Baltz}}\ and\ \bibinfo {author} {\bibfnamefont {W.}~\bibnamefont {Kraut}},\
  }\href {\doibase 10.1103/PhysRevB.23.5590} {\bibfield  {journal} {\bibinfo
  {journal} {Phys. Rev. B}\ }\textbf {\bibinfo {volume} {23}},\ \bibinfo
  {pages} {5590} (\bibinfo {year} {1981})}\BibitemShut {NoStop}%
\bibitem [{\citenamefont {Sipe}\ and\ \citenamefont {Shkrebtii}(2000)}]{Sipe}%
  \BibitemOpen
  \bibfield  {author} {\bibinfo {author} {\bibfnamefont {J.~E.}\ \bibnamefont
  {Sipe}}\ and\ \bibinfo {author} {\bibfnamefont {A.~I.}\ \bibnamefont
  {Shkrebtii}},\ }\href {\doibase 10.1103/PhysRevB.61.5337} {\bibfield
  {journal} {\bibinfo  {journal} {Phys. Rev. B}\ }\textbf {\bibinfo {volume}
  {61}},\ \bibinfo {pages} {5337} (\bibinfo {year} {2000})}\BibitemShut
  {NoStop}%
\bibitem [{\citenamefont {Young}\ and\ \citenamefont
  {Rappe}(2012)}]{Young-Rappe}%
  \BibitemOpen
  \bibfield  {author} {\bibinfo {author} {\bibfnamefont {S.~M.}\ \bibnamefont
  {Young}}\ and\ \bibinfo {author} {\bibfnamefont {A.~M.}\ \bibnamefont
  {Rappe}},\ }\href {\doibase 10.1103/PhysRevLett.109.116601} {\bibfield
  {journal} {\bibinfo  {journal} {Phys. Rev. Lett.}\ }\textbf {\bibinfo
  {volume} {109}},\ \bibinfo {pages} {116601} (\bibinfo {year}
  {2012})}\BibitemShut {NoStop}%
\bibitem [{\citenamefont {Young}\ \emph {et~al.}(2012)\citenamefont {Young},
  \citenamefont {Zheng},\ and\ \citenamefont {Rappe}}]{Young-Zheng-Rappe}%
  \BibitemOpen
  \bibfield  {author} {\bibinfo {author} {\bibfnamefont {S.~M.}\ \bibnamefont
  {Young}}, \bibinfo {author} {\bibfnamefont {F.}~\bibnamefont {Zheng}}, \ and\
  \bibinfo {author} {\bibfnamefont {A.~M.}\ \bibnamefont {Rappe}},\ }\href
  {\doibase 10.1103/PhysRevLett.109.236601} {\bibfield  {journal} {\bibinfo
  {journal} {Phys. Rev. Lett.}\ }\textbf {\bibinfo {volume} {109}},\ \bibinfo
  {pages} {236601} (\bibinfo {year} {2012})}\BibitemShut {NoStop}%
\bibitem [{\citenamefont {Cook}\ \emph {et~al.}(2017)\citenamefont {Cook},
  \citenamefont {Fregoso}, \citenamefont {De~Juan}, \citenamefont {Coh},\ and\
  \citenamefont {Moore}}]{Cook17}%
  \BibitemOpen
  \bibfield  {author} {\bibinfo {author} {\bibfnamefont {A.~M.}\ \bibnamefont
  {Cook}}, \bibinfo {author} {\bibfnamefont {B.~M.}\ \bibnamefont {Fregoso}},
  \bibinfo {author} {\bibfnamefont {F.}~\bibnamefont {De~Juan}}, \bibinfo
  {author} {\bibfnamefont {S.}~\bibnamefont {Coh}}, \ and\ \bibinfo {author}
  {\bibfnamefont {J.~E.}\ \bibnamefont {Moore}},\ }\href@noop {} {\bibfield
  {journal} {\bibinfo  {journal} {Nature communications}\ }\textbf {\bibinfo
  {volume} {8}},\ \bibinfo {pages} {14176} (\bibinfo {year}
  {2017})}\BibitemShut {NoStop}%
\bibitem [{\citenamefont {Morimoto}\ and\ \citenamefont
  {Nagaosa}(2016)}]{Morimoto-Nagaosa16}%
  \BibitemOpen
  \bibfield  {author} {\bibinfo {author} {\bibfnamefont {T.}~\bibnamefont
  {Morimoto}}\ and\ \bibinfo {author} {\bibfnamefont {N.}~\bibnamefont
  {Nagaosa}},\ }\href {\doibase 10.1126/sciadv.1501524} {\bibfield  {journal}
  {\bibinfo  {journal} {Science Advances}\ }\textbf {\bibinfo {volume} {2}},\
  \bibinfo {pages} {e1501524} (\bibinfo {year} {2016})}\BibitemShut {NoStop}%
\bibitem [{\citenamefont {Nie}\ \emph {et~al.}(2015)\citenamefont {Nie},
  \citenamefont {Tsai}, \citenamefont {Asadpour}, \citenamefont {Blancon},
  \citenamefont {Neukirch}, \citenamefont {Gupta}, \citenamefont {Crochet},
  \citenamefont {Chhowalla}, \citenamefont {Tretiak}, \citenamefont {Alam},
  \citenamefont {Wang},\ and\ \citenamefont {Mohite}}]{Nie}%
  \BibitemOpen
  \bibfield  {author} {\bibinfo {author} {\bibfnamefont {W.}~\bibnamefont
  {Nie}}, \bibinfo {author} {\bibfnamefont {H.}~\bibnamefont {Tsai}}, \bibinfo
  {author} {\bibfnamefont {R.}~\bibnamefont {Asadpour}}, \bibinfo {author}
  {\bibfnamefont {J.-C.}\ \bibnamefont {Blancon}}, \bibinfo {author}
  {\bibfnamefont {A.~J.}\ \bibnamefont {Neukirch}}, \bibinfo {author}
  {\bibfnamefont {G.}~\bibnamefont {Gupta}}, \bibinfo {author} {\bibfnamefont
  {J.~J.}\ \bibnamefont {Crochet}}, \bibinfo {author} {\bibfnamefont
  {M.}~\bibnamefont {Chhowalla}}, \bibinfo {author} {\bibfnamefont
  {S.}~\bibnamefont {Tretiak}}, \bibinfo {author} {\bibfnamefont {M.~A.}\
  \bibnamefont {Alam}}, \bibinfo {author} {\bibfnamefont {H.-L.}\ \bibnamefont
  {Wang}}, \ and\ \bibinfo {author} {\bibfnamefont {A.~D.}\ \bibnamefont
  {Mohite}},\ }\href {\doibase 10.1126/science.aaa0472} {\bibfield  {journal}
  {\bibinfo  {journal} {Science}\ }\textbf {\bibinfo {volume} {347}},\ \bibinfo
  {pages} {522} (\bibinfo {year} {2015})}\BibitemShut {NoStop}%
\bibitem [{\citenamefont {Shi}\ \emph {et~al.}(2015)\citenamefont {Shi},
  \citenamefont {Adinolfi}, \citenamefont {Comin}, \citenamefont {Yuan},
  \citenamefont {Alarousu}, \citenamefont {Buin}, \citenamefont {Chen},
  \citenamefont {Hoogland}, \citenamefont {Rothenberger}, \citenamefont
  {Katsiev}, \citenamefont {Losovyj}, \citenamefont {Zhang}, \citenamefont
  {Dowben}, \citenamefont {Mohammed}, \citenamefont {Sargent},\ and\
  \citenamefont {Bakr}}]{Shi}%
  \BibitemOpen
  \bibfield  {author} {\bibinfo {author} {\bibfnamefont {D.}~\bibnamefont
  {Shi}}, \bibinfo {author} {\bibfnamefont {V.}~\bibnamefont {Adinolfi}},
  \bibinfo {author} {\bibfnamefont {R.}~\bibnamefont {Comin}}, \bibinfo
  {author} {\bibfnamefont {M.}~\bibnamefont {Yuan}}, \bibinfo {author}
  {\bibfnamefont {E.}~\bibnamefont {Alarousu}}, \bibinfo {author}
  {\bibfnamefont {A.}~\bibnamefont {Buin}}, \bibinfo {author} {\bibfnamefont
  {Y.}~\bibnamefont {Chen}}, \bibinfo {author} {\bibfnamefont {S.}~\bibnamefont
  {Hoogland}}, \bibinfo {author} {\bibfnamefont {A.}~\bibnamefont
  {Rothenberger}}, \bibinfo {author} {\bibfnamefont {K.}~\bibnamefont
  {Katsiev}}, \bibinfo {author} {\bibfnamefont {Y.}~\bibnamefont {Losovyj}},
  \bibinfo {author} {\bibfnamefont {X.}~\bibnamefont {Zhang}}, \bibinfo
  {author} {\bibfnamefont {P.~A.}\ \bibnamefont {Dowben}}, \bibinfo {author}
  {\bibfnamefont {O.~F.}\ \bibnamefont {Mohammed}}, \bibinfo {author}
  {\bibfnamefont {E.~H.}\ \bibnamefont {Sargent}}, \ and\ \bibinfo {author}
  {\bibfnamefont {O.~M.}\ \bibnamefont {Bakr}},\ }\href {\doibase
  10.1126/science.aaa2725} {\bibfield  {journal} {\bibinfo  {journal}
  {Science}\ }\textbf {\bibinfo {volume} {347}},\ \bibinfo {pages} {519}
  (\bibinfo {year} {2015})}\BibitemShut {NoStop}%
\bibitem [{\citenamefont {de~Quilettes}\ \emph {et~al.}(2015)\citenamefont
  {de~Quilettes}, \citenamefont {Vorpahl}, \citenamefont {Stranks},
  \citenamefont {Nagaoka}, \citenamefont {Eperon}, \citenamefont {Ziffer},
  \citenamefont {Snaith},\ and\ \citenamefont {Ginger}}]{deQuilettes}%
  \BibitemOpen
  \bibfield  {author} {\bibinfo {author} {\bibfnamefont {D.~W.}\ \bibnamefont
  {de~Quilettes}}, \bibinfo {author} {\bibfnamefont {S.~M.}\ \bibnamefont
  {Vorpahl}}, \bibinfo {author} {\bibfnamefont {S.~D.}\ \bibnamefont
  {Stranks}}, \bibinfo {author} {\bibfnamefont {H.}~\bibnamefont {Nagaoka}},
  \bibinfo {author} {\bibfnamefont {G.~E.}\ \bibnamefont {Eperon}}, \bibinfo
  {author} {\bibfnamefont {M.~E.}\ \bibnamefont {Ziffer}}, \bibinfo {author}
  {\bibfnamefont {H.~J.}\ \bibnamefont {Snaith}}, \ and\ \bibinfo {author}
  {\bibfnamefont {D.~S.}\ \bibnamefont {Ginger}},\ }\href {\doibase
  10.1126/science.aaa5333} {\bibfield  {journal} {\bibinfo  {journal}
  {Science}\ }\textbf {\bibinfo {volume} {348}},\ \bibinfo {pages} {683}
  (\bibinfo {year} {2015})}\BibitemShut {NoStop}%
\bibitem [{\citenamefont {Bhatnagar}\ \emph {et~al.}(2013)\citenamefont
  {Bhatnagar}, \citenamefont {Chaudhuri}, \citenamefont {Kim}, \citenamefont
  {Hesse},\ and\ \citenamefont {Alexe}}]{Bhatnagar}%
  \BibitemOpen
  \bibfield  {author} {\bibinfo {author} {\bibfnamefont {A.}~\bibnamefont
  {Bhatnagar}}, \bibinfo {author} {\bibfnamefont {A.~R.}\ \bibnamefont
  {Chaudhuri}}, \bibinfo {author} {\bibfnamefont {Y.~H.}\ \bibnamefont {Kim}},
  \bibinfo {author} {\bibfnamefont {D.}~\bibnamefont {Hesse}}, \ and\ \bibinfo
  {author} {\bibfnamefont {M.}~\bibnamefont {Alexe}},\ }\href {\doibase
  {10.1038/ncomms3835}} {\bibfield  {journal} {\bibinfo  {journal} {{Nat.
  Commun.}}\ }\textbf {\bibinfo {volume} {{4}}},\ \bibinfo {pages} {{2835}}
  (\bibinfo {year} {{2013}})}\BibitemShut {NoStop}%
\bibitem [{\citenamefont {Nagaosa}\ and\ \citenamefont
  {Morimoto}(2017)}]{Nagaosa-Morimoto17}%
  \BibitemOpen
  \bibfield  {author} {\bibinfo {author} {\bibfnamefont {N.}~\bibnamefont
  {Nagaosa}}\ and\ \bibinfo {author} {\bibfnamefont {T.}~\bibnamefont
  {Morimoto}},\ }\href {\doibase 10.1002/adma.201603345} {\bibfield  {journal}
  {\bibinfo  {journal} {Advanced Materials}\ }\textbf {\bibinfo {volume}
  {29}},\ \bibinfo {pages} {1603345} (\bibinfo {year} {2017})}\BibitemShut
  {NoStop}%
\bibitem [{\citenamefont {Resta}(1994)}]{Resta}%
  \BibitemOpen
  \bibfield  {author} {\bibinfo {author} {\bibfnamefont {R.}~\bibnamefont
  {Resta}},\ }\href {\doibase 10.1103/RevModPhys.66.899} {\bibfield  {journal}
  {\bibinfo  {journal} {Rev. Mod. Phys.}\ }\textbf {\bibinfo {volume} {66}},\
  \bibinfo {pages} {899} (\bibinfo {year} {1994})}\BibitemShut {NoStop}%
\bibitem [{\citenamefont {{Sotome}}\ \emph {et~al.}(2018)\citenamefont
  {{Sotome}}, \citenamefont {{Nakamura}}, \citenamefont {{Fujioka}},
  \citenamefont {{Ogino}}, \citenamefont {{Kaneko}}, \citenamefont
  {{Morimoto}}, \citenamefont {{Zhang}}, \citenamefont {{Kawasaki}},
  \citenamefont {{Nagaosa}}, \citenamefont {{Tokura}},\ and\ \citenamefont
  {{Ogawa}}}]{Sotome18}%
  \BibitemOpen
  \bibfield  {author} {\bibinfo {author} {\bibfnamefont {M.}~\bibnamefont
  {{Sotome}}}, \bibinfo {author} {\bibfnamefont {M.}~\bibnamefont
  {{Nakamura}}}, \bibinfo {author} {\bibfnamefont {J.}~\bibnamefont
  {{Fujioka}}}, \bibinfo {author} {\bibfnamefont {M.}~\bibnamefont {{Ogino}}},
  \bibinfo {author} {\bibfnamefont {Y.}~\bibnamefont {{Kaneko}}}, \bibinfo
  {author} {\bibfnamefont {T.}~\bibnamefont {{Morimoto}}}, \bibinfo {author}
  {\bibfnamefont {Y.}~\bibnamefont {{Zhang}}}, \bibinfo {author} {\bibfnamefont
  {M.}~\bibnamefont {{Kawasaki}}}, \bibinfo {author} {\bibfnamefont
  {N.}~\bibnamefont {{Nagaosa}}}, \bibinfo {author} {\bibfnamefont
  {Y.}~\bibnamefont {{Tokura}}}, \ and\ \bibinfo {author} {\bibfnamefont
  {N.}~\bibnamefont {{Ogawa}}},\ }\href@noop {} {\bibfield  {journal} {\bibinfo
   {journal} {arXiv:1801.10297}\ } (\bibinfo {year} {2018})}\BibitemShut
  {NoStop}%
\bibitem [{\citenamefont {Onoda}\ \emph {et~al.}(2006)\citenamefont {Onoda},
  \citenamefont {Sugimoto},\ and\ \citenamefont {Nagaosa}}]{Onoda06}%
  \BibitemOpen
  \bibfield  {author} {\bibinfo {author} {\bibfnamefont {S.}~\bibnamefont
  {Onoda}}, \bibinfo {author} {\bibfnamefont {N.}~\bibnamefont {Sugimoto}}, \
  and\ \bibinfo {author} {\bibfnamefont {N.}~\bibnamefont {Nagaosa}},\ }\href
  {\doibase 10.1143/PTP.116.61} {\bibfield  {journal} {\bibinfo  {journal}
  {Progress of Theoretical Physics}\ }\textbf {\bibinfo {volume} {116}},\
  \bibinfo {pages} {61} (\bibinfo {year} {2006})}\BibitemShut {NoStop}%
\bibitem [{\citenamefont {Sugimoto}\ \emph {et~al.}(2008)\citenamefont
  {Sugimoto}, \citenamefont {Onoda},\ and\ \citenamefont
  {Nagaosa}}]{Sugimoto08}%
  \BibitemOpen
  \bibfield  {author} {\bibinfo {author} {\bibfnamefont {N.}~\bibnamefont
  {Sugimoto}}, \bibinfo {author} {\bibfnamefont {S.}~\bibnamefont {Onoda}}, \
  and\ \bibinfo {author} {\bibfnamefont {N.}~\bibnamefont {Nagaosa}},\ }\href
  {\doibase 10.1103/PhysRevB.78.155104} {\bibfield  {journal} {\bibinfo
  {journal} {Phys. Rev. B}\ }\textbf {\bibinfo {volume} {78}},\ \bibinfo
  {pages} {155104} (\bibinfo {year} {2008})}\BibitemShut {NoStop}%
\bibitem [{\citenamefont {Morimoto}\ and\ \citenamefont
  {Nagaosa}(2018)}]{Morimoto18}%
  \BibitemOpen
  \bibfield  {author} {\bibinfo {author} {\bibfnamefont {T.}~\bibnamefont
  {Morimoto}}\ and\ \bibinfo {author} {\bibfnamefont {N.}~\bibnamefont
  {Nagaosa}},\ }\href@noop {} {\bibfield  {journal} {\bibinfo  {journal} {Sci.
  Rep.}\ }\textbf {\bibinfo {volume} {8}},\ \bibinfo {pages} {2973} (\bibinfo
  {year} {2018})}\BibitemShut {NoStop}%
\bibitem [{\citenamefont {Imry}(2002)}]{Imry}%
  \BibitemOpen
  \bibfield  {author} {\bibinfo {author} {\bibfnamefont {Y.}~\bibnamefont
  {Imry}},\ }\href {https://books.google.com/books?id=ZyjW37iGhaQC} {\emph
  {\bibinfo {title} {Introduction to Mesoscopic Physics}}},\ Mesoscopic physics
  and nanotechnology\ (\bibinfo  {publisher} {Oxford University Press,
  Oxford},\ \bibinfo {year} {2002})\BibitemShut {NoStop}%
\bibitem [{\citenamefont {Lifshitz}\ and\ \citenamefont
  {Pitaevskii}(1980)}]{Lifshitz-Pitaevskii}%
  \BibitemOpen
  \bibfield  {author} {\bibinfo {author} {\bibfnamefont {E.}~\bibnamefont
  {Lifshitz}}\ and\ \bibinfo {author} {\bibfnamefont {L.}~\bibnamefont
  {Pitaevskii}},\ }\href {https://books.google.com/books?id=lgiBDAAAQBAJ}
  {\emph {\bibinfo {title} {Statistical Physics, Part 2: Volume 9}}},\ \bibinfo
  {number} {vol. 9}\ (\bibinfo  {publisher} {Butterworth-Heinemann, Oxford},\
  \bibinfo {year} {1980})\BibitemShut {NoStop}%
\bibitem [{\citenamefont {van~der Marel}(2015)}]{Marel15}%
  \BibitemOpen
  \bibfield  {author} {\bibinfo {author} {\bibfnamefont {D.}~\bibnamefont
  {van~der Marel}},\ }in\ \href@noop {} {\emph {\bibinfo {booktitle} {Strongly
  Correlated Systems}}}\ (\bibinfo  {publisher} {Springer},\ \bibinfo {year}
  {2015})\ pp.\ \bibinfo {pages} {269--296}\BibitemShut {NoStop}%
\bibitem [{\citenamefont {Ando}\ \emph {et~al.}(1998)\citenamefont {Ando},
  \citenamefont {Nakanishi},\ and\ \citenamefont {Saito}}]{Ando98}%
  \BibitemOpen
  \bibfield  {author} {\bibinfo {author} {\bibfnamefont {T.}~\bibnamefont
  {Ando}}, \bibinfo {author} {\bibfnamefont {T.}~\bibnamefont {Nakanishi}}, \
  and\ \bibinfo {author} {\bibfnamefont {R.}~\bibnamefont {Saito}},\ }\href
  {\doibase 10.1143/JPSJ.67.2857} {\bibfield  {journal} {\bibinfo  {journal}
  {J. Phys. Soc. Jpn.}\ }\textbf {\bibinfo {volume} {67}},\ \bibinfo {pages}
  {2857} (\bibinfo {year} {1998})}\BibitemShut {NoStop}%
\bibitem [{\citenamefont {Koshino}\ and\ \citenamefont
  {McCann}(2009)}]{Koshino-McCann09}%
  \BibitemOpen
  \bibfield  {author} {\bibinfo {author} {\bibfnamefont {M.}~\bibnamefont
  {Koshino}}\ and\ \bibinfo {author} {\bibfnamefont {E.}~\bibnamefont
  {McCann}},\ }\href {\doibase 10.1103/PhysRevB.80.165409} {\bibfield
  {journal} {\bibinfo  {journal} {Phys. Rev. B}\ }\textbf {\bibinfo {volume}
  {80}},\ \bibinfo {pages} {165409} (\bibinfo {year} {2009})}\BibitemShut
  {NoStop}%
\bibitem [{\citenamefont {McCann}\ and\ \citenamefont
  {Koshino}(2013)}]{McCann-Koshino13}%
  \BibitemOpen
  \bibfield  {author} {\bibinfo {author} {\bibfnamefont {E.}~\bibnamefont
  {McCann}}\ and\ \bibinfo {author} {\bibfnamefont {M.}~\bibnamefont
  {Koshino}},\ }\href {http://stacks.iop.org/0034-4885/76/i=5/a=056503}
  {\bibfield  {journal} {\bibinfo  {journal} {Reports on Progress in Physics}\
  }\textbf {\bibinfo {volume} {76}},\ \bibinfo {pages} {056503} (\bibinfo
  {year} {2013})}\BibitemShut {NoStop}%
\bibitem [{\citenamefont {Fu}(2009)}]{Fu09}%
  \BibitemOpen
  \bibfield  {author} {\bibinfo {author} {\bibfnamefont {L.}~\bibnamefont
  {Fu}},\ }\href {\doibase 10.1103/PhysRevLett.103.266801} {\bibfield
  {journal} {\bibinfo  {journal} {Phys. Rev. Lett.}\ }\textbf {\bibinfo
  {volume} {103}},\ \bibinfo {pages} {266801} (\bibinfo {year}
  {2009})}\BibitemShut {NoStop}%
\bibitem [{Note1()}]{Note1}%
  \BibitemOpen
  \bibinfo {note} {We note that the nonlinear conductivity in TI surface states
  show a different $B$ dependence ($\propto B^{\protect \frac 3 2}$ from $\ell
  \propto 1/\protect \sqrt {B}$) compared to graphene ($\propto B$). This is a
  consequence of different forms of trigonal warping ($\propto p^3$ in TI and
  $\propto p^2$ in graphene).}\BibitemShut {Stop}%
\bibitem [{\citenamefont {Korm\'anyos}\ \emph {et~al.}(2013)\citenamefont
  {Korm\'anyos}, \citenamefont {Z\'olyomi}, \citenamefont {Drummond},
  \citenamefont {Rakyta}, \citenamefont {Burkard},\ and\ \citenamefont
  {Fal'ko}}]{Kormanyos13}%
  \BibitemOpen
  \bibfield  {author} {\bibinfo {author} {\bibfnamefont {A.}~\bibnamefont
  {Korm\'anyos}}, \bibinfo {author} {\bibfnamefont {V.}~\bibnamefont
  {Z\'olyomi}}, \bibinfo {author} {\bibfnamefont {N.~D.}\ \bibnamefont
  {Drummond}}, \bibinfo {author} {\bibfnamefont {P.}~\bibnamefont {Rakyta}},
  \bibinfo {author} {\bibfnamefont {G.}~\bibnamefont {Burkard}}, \ and\
  \bibinfo {author} {\bibfnamefont {V.~I.}\ \bibnamefont {Fal'ko}},\ }\href
  {\doibase 10.1103/PhysRevB.88.045416} {\bibfield  {journal} {\bibinfo
  {journal} {Phys. Rev. B}\ }\textbf {\bibinfo {volume} {88}},\ \bibinfo
  {pages} {045416} (\bibinfo {year} {2013})}\BibitemShut {NoStop}%
\bibitem [{\citenamefont {Rangel}\ \emph {et~al.}(2017)\citenamefont {Rangel},
  \citenamefont {Fregoso}, \citenamefont {Mendoza}, \citenamefont {Morimoto},
  \citenamefont {Moore},\ and\ \citenamefont {Neaton}}]{Rangel17}%
  \BibitemOpen
  \bibfield  {author} {\bibinfo {author} {\bibfnamefont {T.}~\bibnamefont
  {Rangel}}, \bibinfo {author} {\bibfnamefont {B.~M.}\ \bibnamefont {Fregoso}},
  \bibinfo {author} {\bibfnamefont {B.~S.}\ \bibnamefont {Mendoza}}, \bibinfo
  {author} {\bibfnamefont {T.}~\bibnamefont {Morimoto}}, \bibinfo {author}
  {\bibfnamefont {J.~E.}\ \bibnamefont {Moore}}, \ and\ \bibinfo {author}
  {\bibfnamefont {J.~B.}\ \bibnamefont {Neaton}},\ }\href {\doibase
  10.1103/PhysRevLett.119.067402} {\bibfield  {journal} {\bibinfo  {journal}
  {Phys. Rev. Lett.}\ }\textbf {\bibinfo {volume} {119}},\ \bibinfo {pages}
  {067402} (\bibinfo {year} {2017})}\BibitemShut {NoStop}%
\bibitem [{\citenamefont {Okada}\ \emph {et~al.}(2016)\citenamefont {Okada},
  \citenamefont {Ogawa}, \citenamefont {Yoshimi}, \citenamefont {Tsukazaki},
  \citenamefont {Takahashi}, \citenamefont {Kawasaki},\ and\ \citenamefont
  {Tokura}}]{Okada16}%
  \BibitemOpen
  \bibfield  {author} {\bibinfo {author} {\bibfnamefont {K.~N.}\ \bibnamefont
  {Okada}}, \bibinfo {author} {\bibfnamefont {N.}~\bibnamefont {Ogawa}},
  \bibinfo {author} {\bibfnamefont {R.}~\bibnamefont {Yoshimi}}, \bibinfo
  {author} {\bibfnamefont {A.}~\bibnamefont {Tsukazaki}}, \bibinfo {author}
  {\bibfnamefont {K.~S.}\ \bibnamefont {Takahashi}}, \bibinfo {author}
  {\bibfnamefont {M.}~\bibnamefont {Kawasaki}}, \ and\ \bibinfo {author}
  {\bibfnamefont {Y.}~\bibnamefont {Tokura}},\ }\href {\doibase
  10.1103/PhysRevB.93.081403} {\bibfield  {journal} {\bibinfo  {journal} {Phys.
  Rev. B}\ }\textbf {\bibinfo {volume} {93}},\ \bibinfo {pages} {081403}
  (\bibinfo {year} {2016})}\BibitemShut {NoStop}%
\bibitem [{\citenamefont {Rammer}\ and\ \citenamefont
  {Smith}(1986)}]{Rammer86}%
  \BibitemOpen
  \bibfield  {author} {\bibinfo {author} {\bibfnamefont {J.}~\bibnamefont
  {Rammer}}\ and\ \bibinfo {author} {\bibfnamefont {H.}~\bibnamefont {Smith}},\
  }\href {\doibase 10.1103/RevModPhys.58.323} {\bibfield  {journal} {\bibinfo
  {journal} {Rev. Mod. Phys.}\ }\textbf {\bibinfo {volume} {58}},\ \bibinfo
  {pages} {323} (\bibinfo {year} {1986})}\BibitemShut {NoStop}%
\bibitem [{\citenamefont {Jauho}\ \emph {et~al.}(1994)\citenamefont {Jauho},
  \citenamefont {Wingreen},\ and\ \citenamefont {Meir}}]{Jauho94}%
  \BibitemOpen
  \bibfield  {author} {\bibinfo {author} {\bibfnamefont {A.-P.}\ \bibnamefont
  {Jauho}}, \bibinfo {author} {\bibfnamefont {N.~S.}\ \bibnamefont {Wingreen}},
  \ and\ \bibinfo {author} {\bibfnamefont {Y.}~\bibnamefont {Meir}},\ }\href
  {\doibase 10.1103/PhysRevB.50.5528} {\bibfield  {journal} {\bibinfo
  {journal} {Phys. Rev. B}\ }\textbf {\bibinfo {volume} {50}},\ \bibinfo
  {pages} {5528} (\bibinfo {year} {1994})}\BibitemShut {NoStop}%
\bibitem [{\citenamefont {Aoki}\ \emph {et~al.}(2014)\citenamefont {Aoki},
  \citenamefont {Tsuji}, \citenamefont {Eckstein}, \citenamefont {Kollar},
  \citenamefont {Oka},\ and\ \citenamefont {Werner}}]{Aoki-RMP14}%
  \BibitemOpen
  \bibfield  {author} {\bibinfo {author} {\bibfnamefont {H.}~\bibnamefont
  {Aoki}}, \bibinfo {author} {\bibfnamefont {N.}~\bibnamefont {Tsuji}},
  \bibinfo {author} {\bibfnamefont {M.}~\bibnamefont {Eckstein}}, \bibinfo
  {author} {\bibfnamefont {M.}~\bibnamefont {Kollar}}, \bibinfo {author}
  {\bibfnamefont {T.}~\bibnamefont {Oka}}, \ and\ \bibinfo {author}
  {\bibfnamefont {P.}~\bibnamefont {Werner}},\ }\href {\doibase
  10.1103/RevModPhys.86.779} {\bibfield  {journal} {\bibinfo  {journal} {Rev.
  Mod. Phys.}\ }\textbf {\bibinfo {volume} {86}},\ \bibinfo {pages} {779}
  (\bibinfo {year} {2014})}\BibitemShut {NoStop}%
\bibitem [{\citenamefont {Sturman}\ and\ \citenamefont
  {Fridkin}(1992)}]{Sturman}%
  \BibitemOpen
  \bibfield  {author} {\bibinfo {author} {\bibfnamefont {P.~J.}\ \bibnamefont
  {Sturman}}\ and\ \bibinfo {author} {\bibfnamefont {V.~M.}\ \bibnamefont
  {Fridkin}},\ }\href@noop {} {\emph {\bibinfo {title} {Photovoltaic and
  Photo-refractive Effects in Noncentrosymmetric Materials}}},\ Vol.~\bibinfo
  {volume} {8}\ (\bibinfo  {publisher} {CRC Press, Philadelphia},\ \bibinfo
  {year} {1992})\BibitemShut {NoStop}%
\end{thebibliography}%
